\documentclass[11pt,twocolumn,letterpaper]{article}

\usepackage{enumitem}
\usepackage{multirow}
\usepackage[table,xcdraw]{xcolor}
\usepackage{tabularx}
\usepackage{soul}
\usepackage{xcolor}
\usepackage[framemethod=tikz]{mdframed}
\usepackage{lipsum}
\usepackage{makecell}

\usepackage{flushend,setspace}
\usepackage{algorithm}
\usepackage{algpseudocode}

\usepackage{graphicx}
\usepackage{amsmath}
\usepackage{amssymb}
\usepackage{booktabs}
\usepackage{multirow}
\usepackage[pagebackref,breaklinks,colorlinks]{hyperref}

\usepackage[capitalize]{cleveref}
\usepackage{xcolor}
\crefname{section}{Sec.}{Secs.}
\Crefname{section}{Section}{Sections}
\Crefname{table}{Table}{Tables}
\crefname{table}{Tab.}{Tabs.}

\usepackage{overpic}
\usepackage{enumitem} %< control spacing in itemize/enumerate/...
\usepackage{overpic} %< add raw math symbols to figures
\usepackage{color}
\definecolor{turquoise}{cmyk}{0.65,0,0.1,0.3}
\definecolor{purple}{rgb}{0.65,0,0.65}
\definecolor{dark_green}{rgb}{0, 0.5, 0}
\definecolor{orange}{rgb}{0.8, 0.6, 0.2}
\definecolor{red}{rgb}{0.8, 0.2, 0.2}
\definecolor{darkred}{rgb}{0.6, 0.1, 0.05}
\definecolor{blueish}{rgb}{0.0, 0.3, .6}
\definecolor{light_gray}{rgb}{0.7, 0.7, .7}
\definecolor{pink}{rgb}{1, 0, 1}
\definecolor{greyblue}{rgb}{0.25, 0.25, 1}

\usepackage{blindtext}

\renewcommand{\paragraph}[1]{\vspace{1em}\noindent\textbf{#1}.}
\title{Ranking Loss and Sequestering Learning\\for Reducing Image Search Bias in Histopathology }
\author{Pooria Mazaheri$^1$, Azam Asilian Bidgoli$^1$, Shahryar Rahnamayan$^1$, H.R. Tizhoosh$^2$\\
$^1$NICI Lab, Ontario Tech University, Oshawa, Canada\\
$^2$Rhazes Lab, Mayo Clinic, Rochester, MN USA
}
\date{} % clear date

\begin{document}
\maketitle

\begin{abstract}
Recently, deep learning has started to play an essential role in healthcare applications, including image search in digital pathology.
Despite the recent progress in computer vision, significant issues remain for image searching in histopathology archives. A well-known problem is AI bias and lack of generalization. A more particular shortcoming of deep models is the ignorance toward search functionality. The former affects every model, the latter only search and matching.  
Due to the lack of ranking-based learning, researchers must train models based on the classification error and then use the resultant embedding for image search purposes. Moreover, deep models appear to be prone to internal bias even if using a large image repository of various hospitals. 
This paper proposes two novel ideas to improve image search performance. First, we use a ranking loss function to guide feature extraction toward the matching-oriented nature of the search. By forcing the model to learn the ranking of matched outputs, the representation learning is customized toward image search instead of learning a class label. Second,  we introduce the concept of sequestering learning to enhance the generalization of feature extraction. By excluding the images of the input hospital from the matched outputs, i.e., sequestering the input domain, the institutional bias is reduced. The proposed ideas are implemented and validated through the largest public dataset of whole slide images. The experiments demonstrate superior results compare to \textcolor{black}{the-state-of-art.}
\end{abstract}

\providecommand{\keywords}[1]
{
  \small	
  \textbf{\textit{Keywords---}} #1
}


%########################################

\section{Introduction}
\label{sec:intro}
As digital pathology finds its way into the clinical workflow, there is no suitable provision to access the stored knowledge in archives of whole slides images (WSIs) to make them available to researchers and pathologists for diagnostic and educational purposes. This obstacle in computational pathology is mainly due to the lack of compact and generalizable WSI representations \cite{tizhoosh2018artificial}. Additionally, working with WSIs is faced with various critical challenges, among others, gigapixel image sizes resulting in expensive processing time and lack of expertly annotated images at the pixel level. As a result, most approaches focus on the unsupervised processing of WSI patches to address these limitations. \textcolor{black}{The detailed abbreviations and definitions used in the paper are listed in Table~\ref{abbreviations_table}.}

\begin{table*}[]
\centering
\caption{List of abbreviation used in the paper}
% \scalebox{1.05}{
\begin{tabular}{ll}
\hline
% \cellcolor[HTML]{EFEFEF}            & \cellcolor[HTML]{EFEFEF}       \\
\\
%  \multirow{-2}{*}{\cellcolor[HTML]{EFEFEF}\textbf{Abbreviation}} & \multirow{-2}{*}{\cellcolor[HTML]{EFEFEF}\textbf{Definition}} \\ \hline
\multirow{-2}{*}{\textbf{Abbreviation ~~~~~~~~~  }} & \multirow{-2}{*}{\textbf{Definition}} \\ \hline

 WSI         & Whole Slide Image   \\ \hline 
 TCGA      & The Cancer Genome Atlas    \\ \hline
 DNN      & Deep Neural Network    \\ \hline
 MIL      & Multiple instance learning    \\ \hline
 ACC      & Accuracy    \\ \hline
 CNN      &  Convolutional Neural Network   \\ \hline
 RLF      &  Ranking loss function   \\ \hline
 ISL      &  Instance sequestering learning   \\ \hline

\end{tabular}
\label{abbreviations_table}
\end{table*}

Deep learning has been applied to the challenging task of content-based image search in histopathology, a complex task requiring salient, discriminative, and representative features \cite{dehkharghanian2021selection,adnan2020representation}. These features that should be descriptive of the image content can be extracted by a feature extractor, commonly a deep network \cite{alzu2017content}. For this purpose,  images are fed into a  model, and the output of a specific layer is then utilized for image representation \cite{wan2014deep}. The idea of content-based image search is to compare a feature extracted from a query image to all WSIs in a database in a computationally efficient manner and find the most similar matches \cite{kalra2018content,rasoolijaberi2022multi}. The results can allow researchers to match records of current and past patients and learn from evidently diagnosed and treated cases. Through the information obtained from retrieved similar instances, physicians can hone their initial diagnosis, better understand the patient's prognosis, and eventually devise an accurate final diagnostic interpretation when integrated with ancillary data. However, researchers must train the models based on the classification labels and then use the features to search images. This inconsistency between the model's primary objective (i.e., classification) and the model's downstream task (i.e., search), leads to adverse effects on the results.

In addition to the issue mentioned above,  data bias is another major challenge that should be addressed before integrating deep models into the clinical routine. Bias in  medical studies can be analyzed along with three directions: data-driven, algorithmic, and human bias \cite{norori2021addressing}. Data-driven bias means most fields of human research are heavily biased toward participants and may not be representative of the human population as a whole. Algorithmic bias means that when an algorithm is trained on biased data, it is likely to reinforce patterns from the dominant clues of the data. Finally, human bias may be reflected by human designs of deep models. These types of biases can lead to a lack of generalization and hence misdiagnosis in digital pathology. Due to the propensity of deep learning to overfitting, the model generalization and performance are typically reported in a reserved testing set or evaluated with cross-validation to avoid biased estimates of accuracy \cite{echle2021deep,tommasi2017deeper,shimron2022implicit}. 
However, the overfitting of digital pathology models to site-level characteristics has been incompletely characterized and is infrequently accounted for in the internal validation of models.

Recently,   it has been shown that there are tissue source site-specific patterns of public archives like TCGA images that could be used to classify contributing institutions without any explicit training \cite{dehkharghanian2021biased}; this is clear evidence of existing bias in the images.
It seems that WSIs provided by some institutions may have unique but hidden identifiers that could reveal which institution has shared the data. Consequently, deep models may inadvertently pick up these patterns without explicit training or fine-tuning. The features of a pre-trained DenseNet can almost perfectly detect WSIs that originated from some hospitals \cite{dehkharghanian2021biased}. In addition, a trained model for the classification of cancer subtypes may be able to discover such tissue source site-specific patterns within digital slides to classify cancer types. These findings imply that the trained deep model may have learned to distinguish source site institutions as a form of biased and undesired shortcut to classify cancer types. Generally, this bias affects other chained operations such as segmentation, prediction, and classification. 

To avoid learning the site-specific patterns of institutions, we require to exclude the participation of images from the same institution during the learning. However, the nature of existing classification-based loss functions does not facilitate this idea's development. Hence, we propose a new ranking loss function to train the model based on the similarity among images. 
Then, we sequester the input hospital at the loss function level by proposing a modified version of the ranking loss function.

The paper is structured as follows: Section \ref{Related_Works} briefly covers the related works. Section \ref{Data_Preparations} discusses the EfficientNet as the backbone model and the TCGA repository as our public dataset. Section \ref{Methodology} explains the approach, and experiments and results are reported in Section \ref{Experiments}. Finally, Section \ref{Conclusions} concludes the paper.

%########################
\begin{figure*}
\begin{center}
\includegraphics[width=\textwidth]{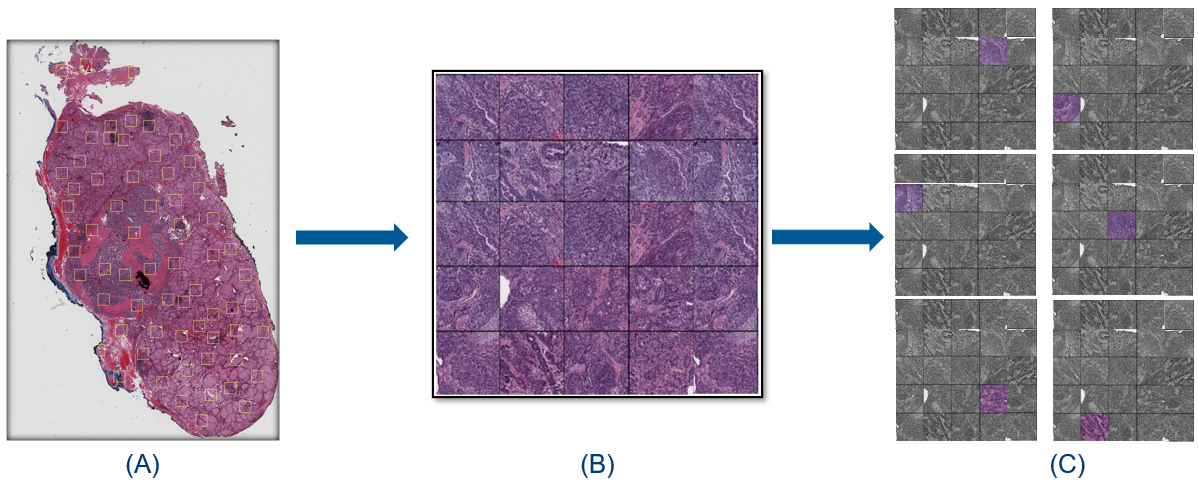}
\end{center}
\caption{
% 
% \textbf{Outline -- }
Extracted patches from a whole slide image (A), created a mosaic from extracted patches (B), and randomly chosen one part from 25 parts for feeding to the network (C).
}
\label{wholepatchselection}
\end{figure*}

\section{Related Works}
\label{Related_Works}

\paragraph{Machine Learning in Digital Pathology}
In recent years, with the continuous advancement of digital pathology  in clinics and laboratories, many research works have proposed to capture the entire glass slide and process it as a digital image \cite{pantanowitz2010digital}. As a sizable number of WSIs is being accumulated, many studies have analyzed WSIs using computer vision based on machine learning to assist in tasks including diagnosis and prognosis \cite{komura2018machine, madabhushi2016image}. However, we are generally faced with lack of effective and expressive techniques for representing WSIs. The characterization of WSIs contains various challenges in terms of image size, color and complexity, and definitiveness of diagnosis at the WSI level. In addition, many efforts are required to annotate a large number of images \cite{adnan2020representation}. All these challenges need more novel ways of representing WSIs \cite{tizhoosh2018representing,fashi2021self}.

\textbf{Content-based Image Search.} In general, image search and retrieval require representative features that are descriptive of the image content. In order to extract these features, researchers employ two types of networks.  Some of them utilize the  pre-trained deep neural networks (DNNs) on images of non-medical objects \cite{komura2018luigi} like the ImageNet dataset \cite{deng2009imagenet}, while others prefer to  fine-tune or train a DNN on pathology images \cite{riasatian2021fine, schaer2019deep}. Narayan et al. \cite{narayan2019similar} used a pre-trained DNN that was trained on natural images to condense an input image into a feature vector.  They adopted a dataset manually annotated by pathologists and reported top-5 scores for patch-based search at various magnification levels. 
In another research, Kalra et al. \cite{kalra2020yottixel} introduced a search engine called \emph{Yottixel} for real-time WSI retrieval in digital pathology. They utilized an unsupervised color-based clustering method to extract a set of images at 20X magnification from each WSI. This set of patches is called the ``mosaic'' of WSI, where approximately 5\% of the tissue specimen is captured. Next, the mosaic is fed to the pre-trained model to extract features. Then, the extracted features are barcoded to create a ``bunch of barcodes'' for fast WSI indexing. The barcoding of gigapixel WSIs enables Yottixel to perform millions of searches in real-time. In another recent paper, Riasatian et al. \cite{riasatian2021fine} proposed image representation for search in digital pathology. They re-trained a model called \emph{KimiaNet} based on DenseNet topology at several configurations using a collected dataset from the TCGA. They adopted a clustering-based approach to select histopathology images at 20X magnification based on a high-cellularity metric. The KimiaNet reported two types of image search, horizontal search and vertical search. In  horizontal search, one searches images across the entire dataset to find the WSI with a similar tumor type to the query WSI among all WSIs. In contrast, in vertical search, one searches images to find similar types of malignancy in an anatomical site, i.e., search for similar WSI with the similar tumor subtype among all WSIs of the same tumor site (same organ).

\textbf{Bias in Digital Medicine.} 
In a recent study, Howard et al. \cite{howard2021impact} reported that the distribution of clinical information in TCGA data, such as survival and gene expression patterns, remarkably differ among samples provided by various clinics and laboratories. They showed that  some models detect source sites instead of predicting prognosis or mutation states. Many works have been conducted to eliminate these site-specific signatures to enhance the validity of histologic image analysis, some through correcting for differences in slide staining between institutions \cite{komura2018machine}. Some of them tried to utilize the proposed methods by  Reinhard et al. and Macenko  \cite{reinhard2001color,macenko2009method} to decrease color variation across images. Other researchers have utilized color augmentation, where the color channels are altered at random during training to prevent a model from learning stain characteristics of a specific site \cite{tellez2018whole,liu2017detecting}.
Most research works of stain-normalization and augmentation techniques have focused on the performance of models in validation sets, rather than elimination of the site-specific signature that may lead to model bias \cite{tellez2019quantifying,anghel2019high}.
In addition,  bias may exist in any type of medical images. 
DeGrave et al. \cite{degrave2021ai} showed that the trained models on radiographic images are more likely to learn medically irrelevant shortcuts, usually attributable to biases in data acquisition, instead of the actual underlying pathology.

In \cite{dehkharghanian2021biased}, the authors revealed that an approach may be subject to bias if the feature extractor is trained on specific institution datasets and potential \emph{hidden} biases are not accounted for. The factors such as  scanner configuration and noise, stain variation and artifacts, and source site patient demographics are more likely potential reasons   for the observed biases \cite{dehkharghanian2021biased}. 

The purpose of this work is not to discover the source of bias. We propose two novel ideas to reduce bias and improve the generalization of deep features for image search. We extend the loss function to consider ranking of matched images. As well, we exclude the images belonging to the input hospital from majority voting among retrieved cases. 

%###############################
\section{Data \& Model Preparations}
\label{Data_Preparations}
\textbf{Image Dataset.} The Cancer Genome Atlas (TCGA) is a joint effort of the National Cancer Institute (NCI) and the National Human Genome Research Institute (NHGRI) that molecularly characterizes tissue samples of more than 11,000 patients along with the primary diagnosis. It  contains histopathology whole slide images (WSIs) of 32 cancer subtypes across 25 primary sites. The WSIs in the TCGA dataset are associated with metadata such as primary diagnosis, tumor stage, and medical center. This repository is used in various deep learning researches that follow different goals such as cancer type classification \cite{liao2020deep,iizuka2020deep,chen2020classification,tabibu2019pan}, somatic mutation prediction \cite{jang2020prediction,coudray2018classification,liao2020deep}, tissue segmentation \cite{de2018automatic,martino2020deep}, and survival estimation \cite{wulczyn2020deep,tabibu2019pan}.

\textcolor{black}{There are many researches that focus on the TCGA \cite{shao2021transmil,liao2020deep}  and implement  and propose different models on it such as deep semi-supervised learning \cite{ge2020deep}. Liao et al. \cite{liao2020deep}. constructed a CNN-based platform using WSIs of hematoxylin and eosin stained digital slides obtained from the TCGA dataset to prediction of somatic mutation. In another research, Shao et al. \cite{shao2021transmil} proposed a new framework called correlated Multiple instance learning (MIL). MIL is considered a powerful tool to address the weakly supervised classification in WSIs based pathology diagnosis. However, the existing MIL methods are usually based on identical and independent distribution hypothesis, thus neglecting the correlation among various instances. Therefore, the authors devised a Transformer based MIL that explored both spatial and morphological information and can effectively deal with unbalanced and multiple classification with great visualization and interpretability. They succeed to achieve faster convergence and better performance over the TCGA dataset.}

\textcolor{black}{\textbf{Deep Topology Backbone.} 
During the last years, developing the convolutional neural network has attracted the attention of many researchers \cite{jalali2019optimal}. According to the recent success of CNNs and amazing architecture of human spinal cord Kabir et al \cite{kabir2022spinalnet} proposed a new DNN architecture with gradual inputs called SpinalNet to achieve higher accuracy with fewer computations. In this network, they split each layer into three splits that lead to reducing the number of weights. In addition, they illustrated the effectiveness of their proposed model on several benchmark datasets leading to the enhancement of the classification accuracy and regression error.
}
Researchers develop convolutional neural networks (CNNs) at a fixed resource cost and then scale up the network to obtain better accuracy. In order to scale the model, they arbitrarily increase depth \cite{he2016deep} or width \cite{zagoruyko2016wide} or use a larger input image resolution \cite{luo2018neural}. Although their methods improve accuracy, usually require tedious manual tuning and often yield suboptimal performance. Tan and Le \cite{tan2019efficientnet} proposed a model scaling method to scale up CNNs in a more structured manner to address these problems. Their method uniformly scales each dimension with a fixed set of scaling coefficients. By utilizing this method, AutoML \cite{he2018amc}, Neural Architecture Search (NAS), and the intelligent and controlled expansion of the three dimensions (i.e., resolution, depth, and width) of an artificial neural network, they developed a family of models, called EfficientNets, which pass the-state-of-the-art accuracy with better efficiency. EfficientNet is a CNN architecture that is commonly used in computer vision due to their high performance in feature learning. There are eight CNN models in the EfficientNet family, including EfficientNet-B0$\sim$EfficientNet-BX - the higher index indicates the  model has a larger network size. Later, the EfficientNet models have been commonly used in several studies for medical image processing. Gupta and Manhas \cite{gupta2021improved} in 2021 proposed a model based on EfficientNet-B3 topology that can efficiently detect oral cancer in histopathology images. Sun et al. \cite{sun2020optimized} proposed a model based on the EfficientNet-B6 architecture for histopathologic cancer detection. They  applied different activation functions as well as different optimization algorithms such as stochastic gradient descent to analyze their effects on diagnosis accuracy. In another research, Wang et al. \cite{wang2021boosted} proposed a boosted EfficientNet model to diagnose the presence of cancer cells in the pathological tissue of breast cancers. This model helped them alleviate the small image resolution problem, which frequently occurs in medical imaging.
Therefore, based on the EfficientNet model's ability to diagnose cancer and pass the-state-of-the-art accuracy with better efficiency, we choose EfficientNet-B0 as a base model for our experiments.

\begin{figure*}[ht]
\begin{center}
\includegraphics[width=\textwidth]{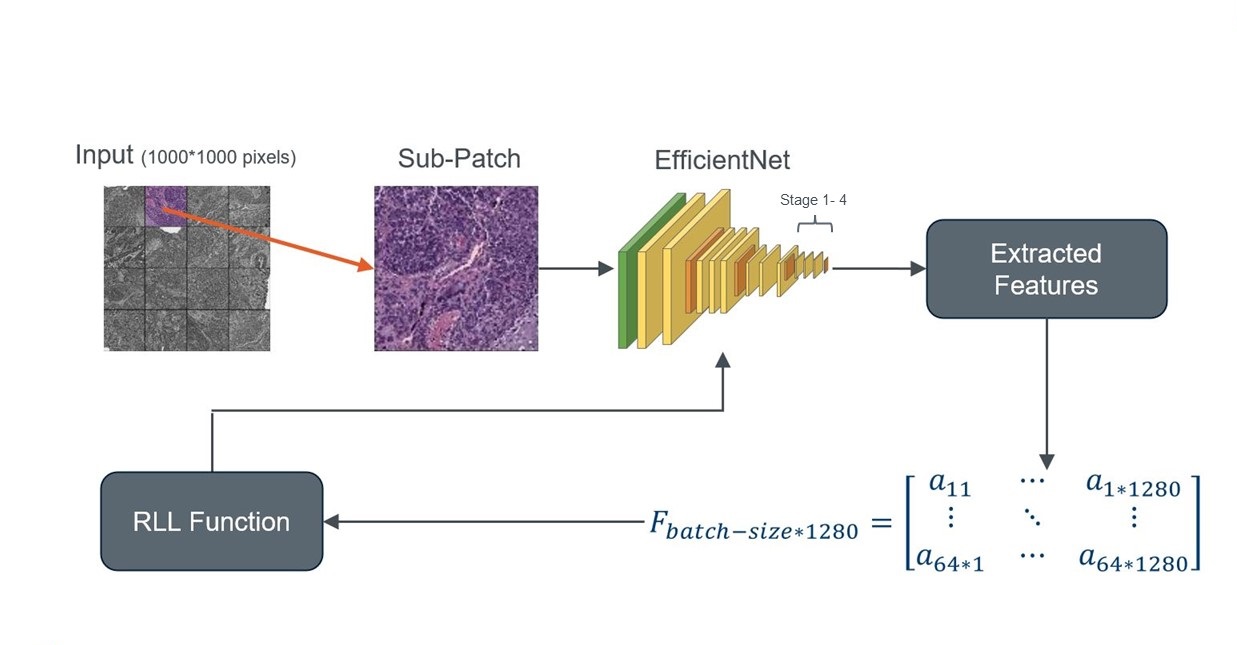}
\end{center}
\caption{
% 
% \textbf{Outline -- }
Overall structure of training process. Firstly, sub-patches are randomly selected from patches and fed to the EfficeintNet network. Then embedded vectors are extracted and passed to the RLF function to calculate the loss value and guide adjusting the weights.
}
\label{InputRLFNetwork}
\end{figure*}

%#############################
\section{Methodology}
\label{Methodology}
The novelty of this research work is based on two ideas to address two independent but often concurring challenges in image search: non-search oriented representations inadequate for similarity matching, and biased representation. The former issue is specific to image search, the latter is a general AI problem; in the following paragraphs, we explain  both challenges in detail.

Image search is one of the promising applications of computerized solutions in digital pathology. It enables pathologists to query patient images or a particular region of interest to find  images with similar histomorphological features within a large WSI archive. However, due to the lack of deep models customized for image search, researchers have to use alternative ways such as the pre-traind classification method instead. Hence, we put forward a new learning guidance mechanism  called \textbf{\emph{ranking loss function}} (RLF) that enables us to train a DNN particularly useful for image search.  

Any loss function may be susceptible to bias. One has to alleviate the bias coming from characteristics of institutions by excluding  the images from the same source site institutions when extracting the features from the WSIs with the same primary diagnosis label. Therefore, we suggest \textbf{\emph{instance sequestering learning}} (ISL) to reduce the bias effect by isolating or hiding from the loss function access. Obviously, ISL can be used along with any  loss function.

\subsection{Ranking Loss Function (RLF)}
As mentioned previously, the basic idea of proposing RLF is to define a loss function based on the ranking of images. Accordingly, similar to what is carried out for image search, the label of an image is predicted  using the labels of similar images. The naive implementation of RLF would require image matching and ranking after each iteration/epoch. However, that would be prohibitively expensive. Hence, we have to establish a more efficient implementation. We create a matrix $\mathbf{F}$ with dimension \emph{batch size} $\times$ {feature vector length}. In the next step, the similarity between each pair of feature vectors belong to $F$ should be calculated. The similarity matrix $\mathbf{S}$ is calculated  as
\begin{equation}
\mathbf{S} = \mathbf{F}\cdot \mathbf{F}
\end{equation}

Since a vector has the highest similarity with itself, the cosine similarity between each vector and itself should not be considered. In order to remove this similarity, we set the diagonal values of matrix to $0$. So we create a modified matrix $\mathbf{S}'$ with diagonal 0 otherwise the same values as in $\mathbf{S}$. In the next step, each row of $\mathbf{S}$ should be normalized to reduce the sensitivity to scale. Then, we build the prediction vector $\mathbf{p}$ of length   \emph{batch size} to be used for loss calculation. In order to predict a label for each input image, the labels of training batch images are used, but with not an equal impact. The most similar feature vector to the query feature vector should exercise the highest effect on predicting the labels. Therefore, we employ a notion of weighted $k$-NN to predict a label for each image. For creating a prediction vector, we use the similarity matrix and actual label vector $\mathbf{l}$. The vector $\mathbf{l}$ includes the real labels of  batch images. To predict a label for image $i$-th image, we use other images that belong to the corresponding batch. Similar images have contribute more to predicting the label for the $i$-the image. Thus, the prediction vector $\mathbf{p}$ is calculated using the label vector, including the real labels (i.e., 0 and 1 in our case), as follows:

\begin{equation}
\mathbf{p} = \mathbf{S}' \cdot \mathbf{l}
\label{Prediction}
\end{equation}

In the last step, the mean square error (MSE) is used as the evaluation metric to obtain the RLF between the predicted labels and actual labels:
\begin{equation}
\textrm{Ranking Loss} = \textrm{MSE}(\mathbf{p} , \mathbf{l}).
\end{equation}

The loss values are passed to the stochastic gradient descent (SGD) for backpropagation and training the weights (see Algorithm.~\ref{Algorithm}). 

Figure.~\ref{InputRLFNetwork} shows the training process for RLF function.

\begin{algorithm}[t]
\caption{ RLF Function Procedure}\label{Algorithm}
\begin{algorithmic}[1]
%\footnotesize
\scriptsize
\State $A  \gets readPatches(filename) $ \Comment{Read patches to fill the training  batch}
\State $F  \gets A $ \Comment{Extracted features from patches}
\State $S  \gets F \cdot F$ \Comment{Calculate similarity matrix}
\State $S'  \gets S $ \Comment{Normalize similarity matrix}
\State $P  \gets S' \times L $ \Comment{Calculate prediction vector}
\State $Ranking Loss \gets MSE(P, L) $ \Comment{Calculate ranking Loss}
\State $SGD  \gets Ranking Loss$ \Comment{Pass loss values to SGD for backpropagation}
\end{algorithmic}
\end{algorithm}

\subsection{Instance Sequestering Learning (ISL)}
The RLF  enables the model to be trained  customized toward image search. Besides, it facilitates customizing the DNN for our second goal, i.e., reducing bias impacts because we can sequester some images from label prediction. In order to prevent the bias effect, ISL excludes the images from the same institutions to contribute in predicting labels. In a sense, ISL adheres to \emph{disclosing conflict of interests} and \emph{recusing} (isolating) the hospital from contributing to majority vote. To be more specific, the images that originated from the same institution should be sequestered so that not be able to vote as part of our $k$-NN based decision making. Accordingly, the DNN is prevented from learning the hospital-identifying patterns (whatever they might be)  and instead it attempts to extract the features representing the patterns of cancer types based on available \emph{external} knowledge. 

Since the similarity value in similarity matrix $\mathbf{S}$ between two images from a specific hospital should be zero, we use a bias label matrix for ISL. The bias label matrix $\mathbf{B}$ consists of rows indicating the feature vector of one patch and the columns representing the hospitals. The value of 1 for cell $\mathbf{B}(i,j)$ shows that  the $i$-th image belongs to the $j$-th hospital.  This matrix should ascertain that two images from the same hospital should not vote for each other during training. In order to create this matrix, we modify the hospital labels to a one-hot vector. 

\begin{figure}[]
\centerline{\includegraphics[width=\columnwidth]{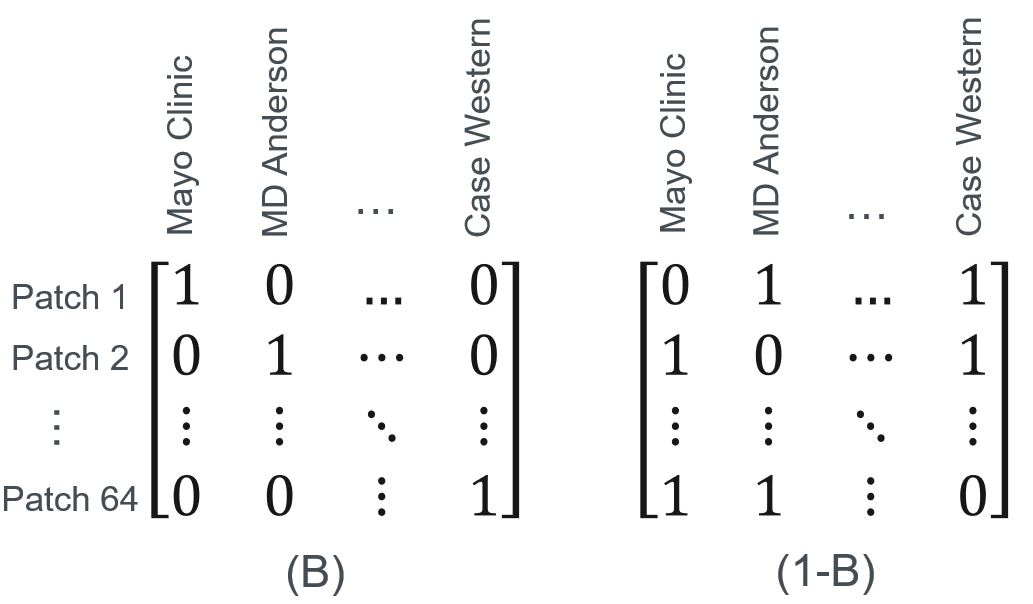}}
\caption{The Bias Label Matrix $\mathbf{B}$ and the inverse bias label
matrix $1-\mathbf{B}$. The columns show the hospitals, and the rows show
the feature vectors. The number of rows is equal to the batch size, and
the indices show the feature vector belongs to a specific hospital
or not.}
\label{BiasLabelVector}
\end{figure}

Figure.~\ref{BiasLabelVector} illustrates an example for matrix $\mathbf{B}$ and $1-\mathbf{B}$ that is used to implement ISL. In order to prevent the voting of same hospital, we flip  the values and create the matrix  $1-\mathbf(B)$. 
Now we use bias label matrix to create the similarity matrix $\mathbf{S}'$:

\begin{equation}
\mathbf{S}' = \mathbf{S} \cdot (1 - \mathbf{B}) 
\label{similarityMatrix}
\end{equation}

\subsection{Architecture \& Training Data}
\textbf{Deep Network --} We utilized a customized \emph{EfficientNet} network \cite{pooria2022custom} to identify the primary diagnosis of histopathology images. For this purpose, as seen in Table~\ref{EfficientNet-Arch}, we added four layers to the EfficientNet topology. Firstly, \emph{batch normalization} (BN) is added to normalize the input of layers by re-centering and re-scaling \cite{ioffe2015batch}. %This helps the model to be faster and more stable and dramatically reduces the number of training epochs required to train the model. 
In addition, dropout is employed as a regularization technique to control overfitting and to discourage learning an unnecessarily high-complex model \cite{srivastava2014dropout}. Moreover, the softmax function and three Fully Connected (FC) layers have been added to accomoadte the classification of cancer subtypes from TCGA (i.e., primary diagnosis). The number of class labels is two (LUAD, LUSC) for the \textbf{lung dataset} and two (LGG, GBM) for the \textbf{brain dataset}. Also, the non-monotonic swish function \cite{ramachandran2017swish} is chosen as an activation function since it is usually more efficient than ReLU \cite{nair2010rectified}.

\begin{table*}[]
\centering
\caption{Added layers to customize the EfficientNet architecture for histopathology image classification. We added BN to normalize the input,  dropout as a regularization technique, softmax function and three FC layers to change the size of last layer to the number of classes.}
\label{EfficientNet-Arch}
\scalebox{0.99}{
\begin{tabular}{|ccccc|}
\hline
% \multicolumn{5}{|c|}{\cellcolor[HTML]{F2F2F2}\textbf{\textit{Additional Layer}}} \\ \hline
\multicolumn{1}{|c|}{} &
  \multicolumn{1}{c|}{} &
  \multicolumn{1}{c|}{} &
  \multicolumn{1}{c|}{} &
   \\

\multicolumn{1}{|c|}{\multirow{-2}{*}{\textbf{\begin{tabular}[c]{@{}c@{}}Stage \\ \textit{i}\end{tabular}}}} &
  \multicolumn{1}{c|}{\multirow{-2}{*}{\textbf{\begin{tabular}[c]{@{}c@{}}Operator \\ \textit{Fi}\end{tabular}}}} &
  \multicolumn{1}{c|}{\multirow{-2}{*}{\textbf{\begin{tabular}[c]{@{}c@{}}Resolution \\ \textit{Hi*Wi}\end{tabular}}}} &
  \multicolumn{1}{c|}{\multirow{-2}{*}{\textbf{\begin{tabular}[c]{@{}c@{}}\#Channels \\ \textit{Ci}\end{tabular}}}} &
  \multirow{-2}{*}{\textbf{\begin{tabular}[c]{@{}c@{}}\#Layers \\ \textit{Li}\end{tabular}}} \\ \hline
% \multicolumn{5}{|c|}{\cellcolor[HTML]{F2F2F2}\textbf{\textit{EfficientNetB0 Architecture, the network baseline}}} \\ \hline
% \multicolumn{5}{|c|}{\cellcolor[HTML]{F2F2F2}\textbf{\textit{Additional Layer}}} \\ \hline
\multicolumn{1}{|c|}{1} &
  \multicolumn{1}{c|}{BN/Dropout} &
  \multicolumn{1}{c|}{7*7} &
  \multicolumn{1}{c|}{1280} &
  1 \\ \hline
\multicolumn{1}{|c|}{2} &
  \multicolumn{1}{c|}{FC/BN/Swish/Dropout} &
  \multicolumn{1}{c|}{1} &
  \multicolumn{1}{c|}{512} &
  1 \\ \hline
\multicolumn{1}{|c|}{3} &
  \multicolumn{1}{c|}{FC/BN/Swish} &
  \multicolumn{1}{c|}{1} &
  \multicolumn{1}{c|}{128} &
  1 \\ \hline
\multicolumn{1}{|c|}{4} &
  \multicolumn{1}{c|}{FC/Softmax} &
  \multicolumn{1}{c|}{1} &
  \multicolumn{1}{c|}{NC} &
  1 \\ \hline
\end{tabular}}
\end{table*}

\textbf{Patch Selection --} Since each WSI contains generally irrelevant pixel information, we need to use a method to remove them. Therefore, a color threshold method was utilized to remove non-tissue regions. Accordingly, the proposed method in  Yottixel was used for patch selection \cite{kalra2020yottixel}. This method assembled a ``mosaic'' of each WSI.
The mosaic is a rather small collection of patches to represent the entire WSI. In essence, Yottixel enables us to represent the entire WSI with a small collection of patches. Then, we removed all patches with low cellularity by taking the top 20 percent of cellularity sorted patches \cite{riasatian2021fine}. Figure.~\ref{wholepatchselection}(A) and  (B) illustrate a sample of histopathology image with extracted patches and a created mosaic  of high cellular patches.

\textbf{Grid strategy for training -- }
The customary purpose followed by researchers in training DNNs for digital pathology is to construct a model to discriminate histopathology images based on the primary diagnosis. For this purpose, a portion of data equal to batch size should be fed to the network for learning parameters and training. However, there are two crucial challenges for development. Due to the large size of patches, high resources and memory are needed for preparing and feeding the data into the network. Additionally, the training process of the network can be time-consuming and take many hours, even days. The second challenge is that the size of the patches (i.e., $1000 \times 1000$ pixels) is different from the size of ImageNet samples that the original EfficientNet is trained on.
% Since the size of images in the ImageNet dataset is $224 * 224$, it can be more efficient for models to feed with the same image size as input.
% In order to address these challenges, two strategies can be suggested.
Therefore, we utilized the Grid method as an image cropping solution to address these challenges.
% The first common approach is to resize (i.e., down-sampling) images  to the size of ImageNet input. Although this strategy can solve the mentioned problem, it can cause losing some prominent information and valuable details for cancer diagnoses. 
% The alternative solution is to crop the images. According to this solution, we propose the Grid method. 
% This method can lead to fast training and reducing the need for expensive processing resources. Accordingly, we used $8$ GB memory with $1$ GPU, while the required resources for training the-state-of-art network has reportedly been $128$ GB memory with $4$ GPUs.

For preparing training datasets, firstly, we mesh each image into $25$ sub-patches with the size of $224\times 224$ pixels. Next, we randomly choose one of them and feed it to the network in each epoch during the training phase (Figure.~\ref{wholepatchselection}(C)). At the next step, each randomly selected part is fed into a pre-trained EfficientNet for feature extraction. Finally, the extracted features vector with dimension 1280 is passed to the RLF function for calculating loss value.
This method can cover most parts of the original patches to feed the network since it chooses many times (i.e., epoch size) from 25 sub-patches. %Fig.~\ref{wholepatchselection} shows the process of preparing inputs for the model.

%###########################

% \input{tab/sota}
% \input{tab/ablations}
\section{Experiments}
\label{Experiments}
We conducted three series of experiments to assess the trained model, with and without considering the effect of bias on Lung and Brain datasets from the TCGA repository. We used $26,021$ Lung patches from $81$ WSIs , and $36,000$ Brain patches from $74$ WSIs. 
We have two main subtypes of Lung cancer, namely Lung Squamous Cell Carcinoma (LUSC) and Lung Adenocarcinoma (LUAD) that account for around $70\%$ of all lung cancers \cite{graham2018classification}. Classifying these two main subtypes is a crucial step to building computerized decision support and triaging systems. The Brain cancer dataset contains two subtypes, namely  Lower-Grade Glioma (LGG) and Glioblastoma multiforme (GBM).

We run training/fine-tuning in several configurations to validate our RLF and ISL. 
% We used the PyTorch framework to train and test the model on one NVIDIA P100 GPU with 16GB memory.
The model was trained with a batch size of 64 for 20 epochs. The network was initialized with pre-trained weights of the ImageNet \cite{deng2009imagenet} and used the SGD as its optimizer.
In order to evaluate the performance of our method, we use F1-score  \cite{dubey2021decade,kalra2020yottixel} and is calculated as the harmonic mean of the precision and recall:

\begin{equation}
F1 \mbox{-}score = \frac{2 \times Precision \times Recall}{Precision + Recall },
\end{equation}

\noindent where
\begin{equation}
\label{PR}
\mathit{Precision}= \frac{True Positive}{True Positive+False Positive},
\end{equation}
\noindent and

\begin{equation}
\label{RE}
\mathit{Recall}= \frac{True Positive}{True Positive+False Negative}.
\end{equation}\\

\subsection{Evaluation of  Ranking Loss}
In the first series of experiment, we evaluated the performance of RLF (without considering bias) and compared our results with two state-of-the-art models. The first model is the original DenseNet-121 \cite{huang2017densely} trained with $1.2$ million natural images from ImageNet \cite{deng2009imagenet}. The pre-trained DenseNet is utilized to extract feature vectors for histopathology images. The second
model is a state-of-the-art network in digital pathology, called KimiaNet, that is based on the DenseNet-121 architecture but has been trained with a large number of histopathology patches \cite{riasatian2021fine}.

\begin{figure}[ht]
\centerline{\includegraphics[width=\columnwidth]{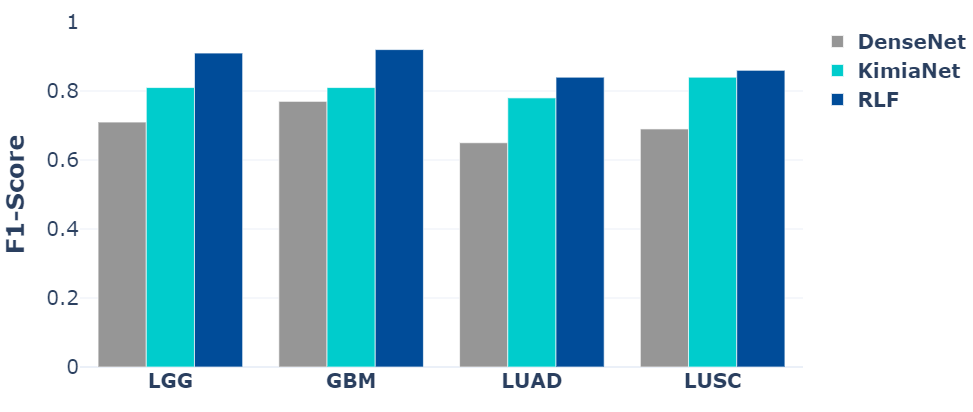}}
\caption{A bar-chart based comparison of the results of the search through representations generated by the DenseNet, KimiaNet, and RLF-trained EfficientNet  on 4 tumor subtypes. The results show that FOR all classes the RLF training performs better than KimiaNet and DenseNet.}
\label{rllkimiadensef1fig}
\end{figure}

Figure.~\ref{rllkimiadensef1fig} represents  the  comparison of image search results  on  two datasets,  obtained  using  EfficientNet with  RLF,  KimiaNet, and DenseNet. In addition, Table~\ref{RLLKimiaDenseTable} provides the F1-score values obtained from three models. As seen, the proposed model achieved  better results for all subtypes of the Lung and Brain cancers. 

The results for the Lung dataset indicate that our proposed
model can improve the accuracy by 6\% and 2\% compared to KimiaNet and 19\% and 17\% compared to DenseNet for two classes, LUAD and LUSC, respectively.  
In another experiment on the Brain dataset, our model performs well on both classes, and the results have significantly enhanced from 71\% and 81\% to 91\%
for LGG and from 77\% and 81\% to 92\% for GBM compared with DenseNet and KimiaNet, respectively.
Thus, the proposed RLF for training the EfficientNet can result in more accurate search image results compared to more complicated DNNs which are trained with the purpose of classification. In other words, it shows utilizing search-based loss functions may perform better than classification-based loss functions for image search.

\begin{figure}[htb]
\centerline{\includegraphics[width=\columnwidth]{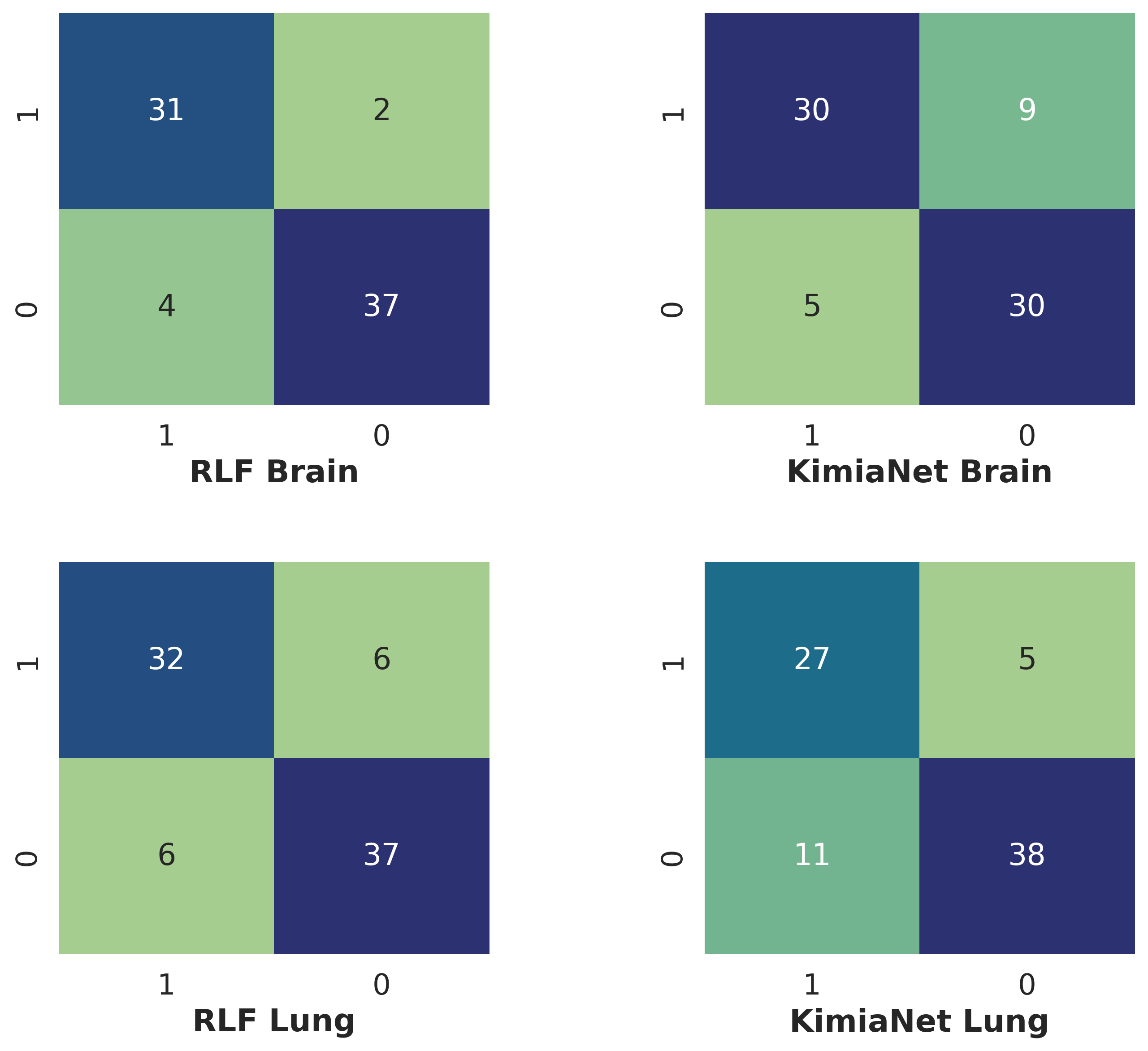}}
\caption{\textcolor{black}{ Confusion matrices of RLF and KimiaNet for the
Lung and Brain dataset.}}
\label{ConfusionMatrixes_Efficientnet&Kimianet}
\end{figure}

\textcolor{black}{In addition to F1-score, in order to show the details of models, the precision and accuracy are calculated for both datasets. 
% As it can be seen in the Table~\ref{RLLvKimiaNetDense}, the accuracy improve near 11\% and 5\% for Brain and Lung datasets respectively.  
% Figure.~\ref{ConfusionMatrixes_Efficientnet&Kimianet} illustrates the confusion matrices of RLF and KimiaNet that show the performance of the models for both Lung and Brain datasets. 
The results for the Lung dataset in the Table~\ref{RLLvKimiaNetDense} indicate that our proposed model can improve the accuracy by 18\% and 5\% compared with DenseNet and Kimianet, respectively. In another experiment on the Brain dataset, our model performs well, and the results have significantly enhanced from 81\% to 92\%. In addition, as can be seen in Table~\ref{RLLvKimiaNetDense}, the precision improve near 3\% and 14\% for the Brain and Lung datasets respectively.Figure.~\ref{ConfusionMatrixes_Efficientnet&Kimianet} illustrates the confusion matrices of RLF and KimiaNet that show the performance of the models for both Lung and Brain datasets. 
}

%%%%%%%%%%%%%%%%%%%%%%%%%%%%%%%%%%%%%%%%%%%%%%%%%%%%%%%%%%%
\begin{table*}[]
\centering
\caption{A comparison  of the result of the image search by  RLF-trained EfficientNet, KimiaNet, and DenseNet that are evaluated on the Lung and Brain tumor types. The results are reported in terms of the F1-score. In addition, the number of training and test samples are given. }
\label{ModelsAccuracy}
\begin{tabular}{|c|c|c|c|c|c|c|}
\hline
\cellcolor[HTML]{EFEFEF}{\color[HTML]{212121} } &

  \cellcolor[HTML]{EFEFEF}{\color[HTML]{212121} } &
  \cellcolor[HTML]{EFEFEF}{\color[HTML]{212121} } &
  \cellcolor[HTML]{EFEFEF}{\color[HTML]{212121} } &
  \cellcolor[HTML]{EFEFEF}{\color[HTML]{212121} } &
  \cellcolor[HTML]{EFEFEF}{\color[HTML]{212121} } &
  \cellcolor[HTML]{EFEFEF}{\color[HTML]{212121} } \\
\multirow{-2}{*}{\cellcolor[HTML]{EFEFEF}{\color[HTML]{212121} \textbf{Dataset}}} &
 
  \multirow{-2}{*}{\cellcolor[HTML]{EFEFEF}{\color[HTML]{212121} \textbf{\#Training}}} &
  \multirow{-2}{*}{\cellcolor[HTML]{EFEFEF}{\color[HTML]{212121} \textbf{\#Test}}} &
  \multirow{-2}{*}{\cellcolor[HTML]{EFEFEF}{\color[HTML]{212121} \textbf{Sub-type}}} &
  \multirow{-2}{*}{\cellcolor[HTML]{EFEFEF}{\color[HTML]{212121} \textbf{DenseNet}}} &
  \multirow{-2}{*}{\cellcolor[HTML]{EFEFEF}{\color[HTML]{212121} \textbf{KimiaNet}}} &
  \multirow{-2}{*}{\cellcolor[HTML]{EFEFEF}{\color[HTML]{212121} \textbf{RLF}}} \\ \hline
 &
  
   &
   &
  LUAD &
  0.65 &
  0.78 &
  \textbf{0.84} \\ \cline{4-7} 
\multirow{-2}{*}{\textbf{Lung}} &
  \multirow{-2}{*}{23321} &
  \multirow{-2}{*}{11535} &
  LUSC &
  0.69 &
  0.84 &
  \textbf{0.86} 
 \\ \hline
 &
   &
   &
  LGG &
  0.71 &
  0.81 &
  \textbf{0.91} \\ \cline{4-7} 
\multirow{-2}{*}{\textbf{Brain}} &
  \multirow{-2}{*}{34629} &
  \multirow{-2}{*}{8068} &
  GBM &
  0.77 &
  0.81 &

  \textbf{0.92} \\ \hline
\end{tabular}
\label{RLLKimiaDenseTable}
\end{table*}

%%%%%%%%%%%%%%%%%%%%%%%%%%%%%%%%%%%%%%%%%%%%%%%%%%%%%%%%%%%%%

In addition to higher accuracy, the image size used for training our model (i.e., $224\times224$) is nearly $20$ times smaller than that of KimiaNet (i.e., $1000\times1000$). The reduced size of inputs makes our model very fast. The proposed model requires just around $20$
minutes for each epoch, whereas the state-of-the-art, KimiaNet, model took approximately 110 minutes for the training
phase on the same platform. Moreover, since the number of parameters was reduced from $7.1$ to $4.3$ millions, nearly $40\%$ reduction, fewer computational resources are needed to train our network. The proposed model demands $8$ GB memory of one GPU while the KimiaNet model used $4$ GPUs
with $128$ GB memory for training. Also, the proposed model is trained for discrimination of cancer subtypes of  a specific organ and is correspondingly capable of training with a smaller dataset, while the KimiaNet is trained on the WSIs from 32 cancer types.
% The proposed model can be trained with a much smaller training set for applications such as image search that require robust and compact representations.
% and capability of training with a smaller datasets.
Table~\ref{RLLvKimiaNetDense}
compares the number of parameters, epoch, batch size, and the
size of inputs for the DenseNet, KimiaNet, and the RFL-trained EfficientNet. 
The time of training for each epoch for DenseNet is not defined because it is not trained.
As it  can be seen, the number of parameters reduced
by $2.8$ millions and the size of patches is almost $20$
times smaller than the state-of-the-art models.

\begin{table*}[hbt]
\centering
\caption{\textcolor{black}{The number of parameters, number of epochs, batch size, and the size of images for different models. It shows the image size that used for proposed model is near 20 times smaller, the EfficientNet with RLF (EN-RLF) is 5.5 times faster than the other models and EN-RLF has better accuracy and precision compare to the KimiaNet and DenseNet for both datasets (P=\#parameters, T=epoch time, PB=precision brain, PL=precision brain, AB=accuracy brain, AL=accuracy lung)}.}
\label{RLLvKimiaNetDense}
\scalebox{0.99}{
\begin{tabular}{|l|l|l|l|l|l|l|l|}
\hline
 Model & Image Size & P & T & PB & PL & AB & AL \\
  \hline
\textbf{DenseNet} & $1000\!\times\!1000$ & 7.1 M   & -   & 0.66 & 0.63 & 0.74 & 0.67    \\ \hline
\textbf{KimiaNet} & $1000\!\times\!1000$ & 7.1 M   & 110 mins & 0.86 & 0.71 & 0.81 & 0.80       \\ \hline
\textbf{ENet-RFL}  & \textbf{$224\!\times\!224$} & \textbf{4.3 M}  & \textbf{20 mins} & \textbf{0.89} & \textbf{0.85} & \textbf{0.92} & \textbf{0.85}\\ \hline
\end{tabular}}
\end{table*}

\subsection{External Validation of ISL}
In the second series of experiments, in order to show the effect of bias reduction on the results, we evaluated the performance of RFL and Instance Sequestering Learning  (ISL) using the external validation. We compared the results from the previous experiment model (i.e., specialized EfficientNet model trained with classification-based loss function) with the proposed loss function and investigated the effect of reducing bias on the results through sequestering. For the purpose of external validation, it was required to separate a portion of data belong to a specific hospital to avoid using them in training and test sets. 

\textbf {Training Phase.} 
%Since the goal of proposed loss functions is to investigate its effect on decreasing the bias,  the external validation is required. 
In order to provide the data for external validation,  we trained as many models as the number of hospitals. Each time, we separated the WSIs of one of the hospitals and trained the model with images of the remaining hospitals. The selected hospital,  therefore, cannot contribute to training and test phases and it can be strictly used for  external validation.  In addition, we collect all hospitals with only one WSI and used them for external validation. We compared the results of  three different types of network. The first type is  the proposed model  based on the classification loss (i.e., proposed EfficientNet trained using classification-based loss function). Then we trained the network using RFL without attempt to sequester data, same as the first experiment. In fact, we intend to evaluate the ranking-based loss function and compare it with the classification-based one. Finally, we trained the model using ISL with RFL and tried to reduce the impact of the bias by sequestering the hospital during training.

\textbf{External validation.} In order to evaluate the performance of the model using instance sequestering, we used the grid method to generate $25$ patches from each image. Then we fed the network with all these patches. As a result, we obtained $25$ feature vectors with the size of $1,280$ from the feature extractor layer for each image. Then, the average of these feature vectors is calculated and considered as the representative for the selected image \cite{azam2022embedding}. As mentioned previously, the extracted feature vectors are used for image search to find the most similar images for a query WSI. However, to evaluate the performance of the search, the primary diagnosis labels are employed. For this purpose, to predict the label of each WSI, we passed each feature vector to the \emph{leave-one-hospital-out} method. Firstly, we calculate the average of all features vectors extracted for patches that belong to one WSI. Then, we used the Euclidean distance to calculate the dissimilarity among the query WSI and the rest of the samples. Next, the 3-NN method is applied to predict the label of query WSI.

\textbf{Image Search on Lung Dataset.}  In order to evaluate the model's performance on the Lung dataset, we fed the unseen images of one specific hospital, not seen in the training/test phase, for external validation. We have done that for $17$ different models and compared the results with the classification  and RFL-based models. Table~\ref{exteranlf1lungtable} reports the F1-score value for these $17$ experiments.

\begin{figure}[!ht]
\centerline{\includegraphics[width=\columnwidth]{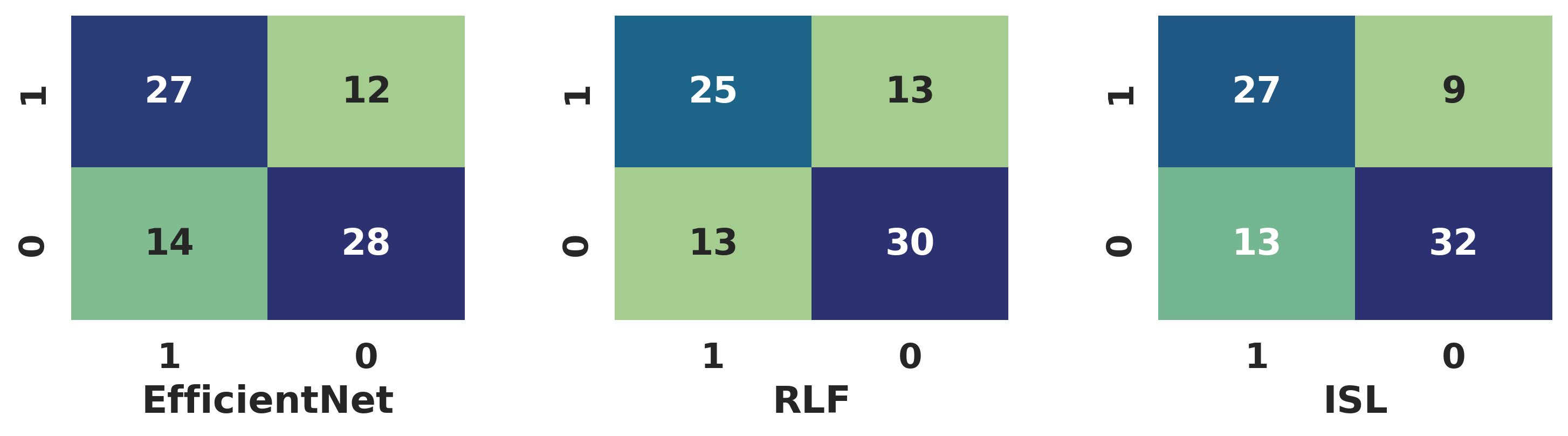}}
\caption{\textcolor{black}{Confusion matrices of EfficientNet, RLF and ISL for the
Lung dataset.}}
\label{ConfusionMatricesLungCancer}
\end{figure}

In these experiments, each time, the train data belonged to one hospital excluded from the training dataset, and the model was evaluated on the test data belonging to the hospital in focus.
% Each model is evaluated using the images of one hospital.
The F1-score of both subtypes are represented separately. In some cases, there is not any sample from one the subtypes, hence the zero F1-score is  obtained. %The results show that SRLL model performances in the $11$ and $6$ experiments are equal to or better than two other methods, the RLL and classification model. 
In most of the cases, squestering learning and ranking loss is more successful in external validation compared to other methods. 
We calculated the total number of correct predictions for each model showing the sequestering and ranking loss can improve the accuracy by nearly $5\%$ compared with two other methods. The last row in  Table~\ref{exteranlf1lungtable} provides the total accuracy from three models. As seen, reducing bias leads to  achieving better results in the Lung dataset in terms of generalization.
\textcolor{black}{Figure.~\ref{ConfusionMatricesLungCancer} illustrates the confusion matrices of EfficientNet, RLF and ISL for external validation on Lung dataset.}

% \usepackage{graphicx}
\begin{table*}[ht]
\centering
\caption{Comparison of F1-Score value for subtypes of Lung cancer that obtained from different models for 17 hospitals as external dataset. The mentioned hospital in each row is excluded from training dataset to utilize its images as an external validation ($n_1=$\#LUAD, $n_2=$ \#LUAC).}
% \resizebox{\textwidth}{!}{%
\begin{tabular}{|l|c|c|c|c|}
\hline
\cellcolor[HTML]{EFEFEF}                               & \cellcolor[HTML]{EFEFEF}                               & \cellcolor[HTML]{EFEFEF}                              & \cellcolor[HTML]{EFEFEF}                                  & \cellcolor[HTML]{EFEFEF}                            \\
\multirow{-2}{*}{\cellcolor[HTML]{EFEFEF}\textbf{Hospital}}   & \multirow{-2}{*}{\cellcolor[HTML]{EFEFEF}\textbf{($n_1$,$n_2$)}}& \multirow{-2}{*}{\cellcolor[HTML]{EFEFEF}\textbf{EfficientNet}} & \multirow{-2}{*}{\cellcolor[HTML]{EFEFEF}\textbf{RLF}} & \multirow{-2}{*}{\cellcolor[HTML]{EFEFEF}\textbf{ISL}} \\ \hline

Intern.   Genomics Consortium             &(8, 8)             & (0.67 ,   0.57)             & (0.59 , 0.53)          & \textbf{(0.78 , 0.71)}\\
Indivumed                                        &(5, 5)            & \textbf{(0.91 , 0.89)}               & (0.83 , 0.75)         & (0.77 , 0.57)\\
Christiana Healthcare                             &(6, 2)             & (0.60 , 0.33)               & \textbf{(0.91 , 0.80)}         & (0.80 , 0.67)\\
Asterand                                          &(1, 8)            & (0.50 , 0.86)                & (0 , 0.94)         & \textbf{(1, 1)} \\
Mayo Clinic Rochester                              &(0, 5)            & (0 , 0.89)                & \textbf{(0 , 1)}        & (0 , 0.75)\\
Washington University-Alabama                &(1, 2)            & \textbf{(1 , 1)}                & (0, 0.80)           & \textbf{(1 , 1)}\\
Roswell Park                                       &(2 ,1)            & (0.67 , \textbf{0.67})               & (0.67 , 0)             & (0, 0.50) \\
Ontario Inst. for Cancer   Research             &(1, 2)            & (0.67 , 0.67)              & (0.67 , 0.67)         & \textbf{(1 , 1)}\\
University of North Carolina                       &(2, 0)           & (0 , 0)               & (0 , 0)         & (0 , 0) \\
Prince Charles Hospital                              &(1, 1)           & (1 , 1)               & (1 , 1)          & (1 , 1)\\
University of Pittsburgh                            &(2, 0)            & ( 0.67 , 0)             & ( 0.67 , 0)        & ( 0.67 , 0) \\
Washington University - Emory                       &(2, 2)            & (0.50 , 0.50)               & (0 , 0.40)       & \textbf{(0.67 , 0.80)}  \\
Memorial Sloan Kettering   Cancer                    &(0, 2)          & ( 0 , 0.67)              & (0 , 0.67)                 & \textbf{(0 , 1)} \\
Washington Uni-  Cleveland Clinic                  &(2, 0)            & (\textbf{0.67} , 0)                   & (0 , \textbf{1})        & (\textbf{0.67} , 0)\\
Thoraxklinik at Uni  Heidelberg                     &(2, 0)             & (0.67 , 0)                   & (0 , 0)         & \textbf{(1 , 0)} \\
Candler                                             &(1,1)           & (0 , 0.67)                      & (0 , 0.67)         & (0 , 0.67) \\
Hospital with one WSI                                &(2,4)           & (0.67 , 0.89)                     & (0.67 , 0.89)          & (0.67 , 0.89)\\ \hline
\textbf{Overall}                            &(38, 43)              & 67.90\%                     & 67.90\%          & \textbf{72.80\%}\\ \hline

\end{tabular}%
%  }

\label{exteranlf1lungtable}
\end{table*}

\textbf{Image Search on Brain Cancer.}  Same as in for search on Lung cancer images, our model received the images of one specific hospital and outputs two classes, GBM or LGG, of Brain tumor subtype. We used this dataset to evaluate $11$ different models and compared the results with other methods. During the training of these models, each time, we excluded one hospital from the training dataset and externally validated the model  belonging to that hospital. 

\begin{figure}[!ht]
\centerline{\includegraphics[width=\columnwidth]{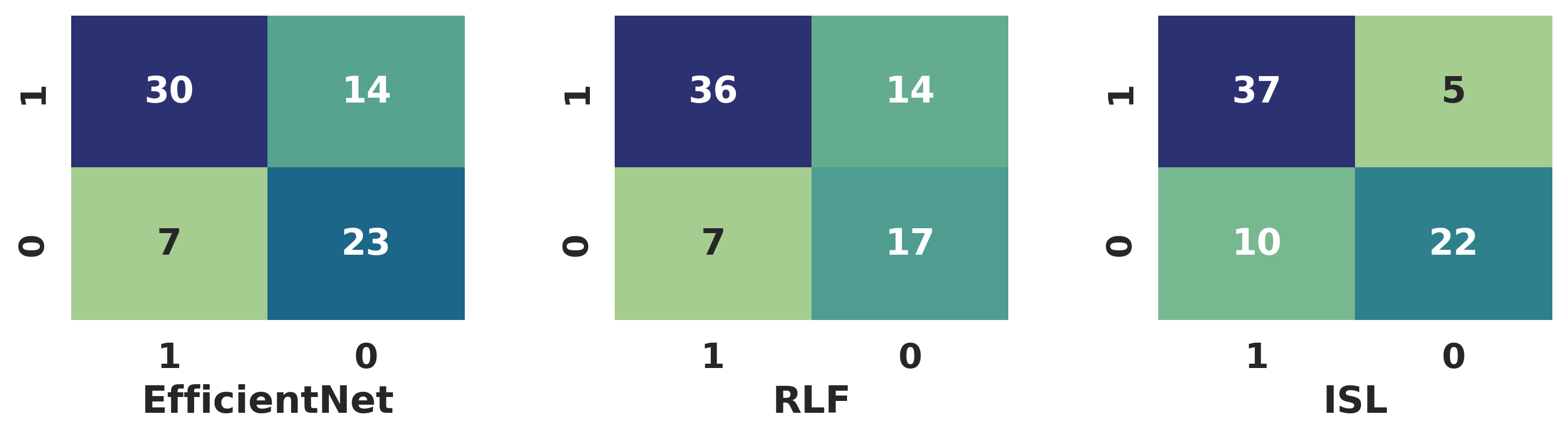}}
\caption{\textcolor{black}{ Confusion matrices of EfficientNet, RLF and ISL for the
Brain dataset.}}
\label{ConfusionMatricesBrainCancer}
\end{figure}

%During 12 experiments on external validation datasets, we were successful on ten experiments and obtained equal or better accuracy.
Table~\ref{exteranlf1braintable} shows the F1-scores for the external validation over $11$ hospitals. Accordingly, the total number of correct predictions for each model was calculated. As seen,  the sequestering with ranking loss can improve the accuracy by nearly $9\%$ compared to two other models. Again, as the number of samples for each tumor subtypes shows, in some hospitals, there is only samples from one of the subtypes and correspondingly, it results in a zero F1-score. 
The last row in Table.~\ref{exteranlf1braintable} shows total accuracy obtained from three models and explains that reducing bias can enhance the results in terms of generalization of the learning. \textcolor{black}{Figure.~\ref{ConfusionMatricesBrainCancer} illustrates the confusion matrices of EfficientNet, RLF and ISL for external validation on Brain dataset.}

% \usepackage{graphicx}
\begin{table*}[]
\centering
\caption{Comparison of F1-Scores for subtypes of Brain cancer that obtained from different models for 12 hospitals as external validations ($n_1=$\#LGG, $n_2=$\#GBM).}
% \resizebox{\textwidth}{!}{%
\begin{tabular}{|l|c|c|c|c|}
\hline
\cellcolor[HTML]{EFEFEF}                               & \cellcolor[HTML]{EFEFEF}                               & \cellcolor[HTML]{EFEFEF}                              & \cellcolor[HTML]{EFEFEF}                               & \cellcolor[HTML]{EFEFEF}                                \\
\multirow{-2}{*}{\cellcolor[HTML]{EFEFEF}\textbf{Hospital}} & \multirow{-2}{*}{\cellcolor[HTML]{EFEFEF}\textbf{($n_1$,$n_2$)}} & \multirow{-2}{*}{\cellcolor[HTML]{EFEFEF}\textbf{EfficientNet}} & \multirow{-2}{*}{\cellcolor[HTML]{EFEFEF}\textbf{RLF}} & \multirow{-2}{*}{\cellcolor[HTML]{EFEFEF}\textbf{ISL}} \\ \hline
University of Sao Paulo         & (0, 2)     & (0 ,   1)                & (0 , 1)                        & (0 , 1) \\
Henry Ford Hospital             & (6, 9)    & (0.57 , 0.63)                    & {(\textbf{0.62} , 0.57)}               & (0.50 , \textbf{0.67})\\
Case   Western - St Joes        & (0, 6)     & (0 , 0.80)                  & \textbf{(0 , 1)}                       & \textbf{(0 , 1)}\\
% Duke University                 & (0, 1)     & (0 , 1)                   &  (0 , 1)                  & (0 , 1)\\
Dept of Neurosurgery at Uof H   & (0, 6)     & (0 , 0.67)                   & \textbf{(0, 0.80)}                       & \textbf{(0 , 0.80)} \\
University of Florida           & (4, 3)     & (0.75 , 0.67)                 & (0.75 , 00.67)                        & \textbf{(0.86 , 0.86)} \\
Mayo Clinic                     & (0, 2)     & \textbf{(0 , 1)}                  & (0 , 0.50)                        & \textbf{(0 , 1)} \\
MD Anderson                     & (7, 0)     & (0.60 , 0)                    & \textbf{(1 , 1)}                    & (0.73 , 00) \\
UCSF & (6, 1)  & (0.80 , 0.50)             & (0.80 , 0.50)                      & (0.80 , 0.50)  \\
Case Western                    & (8, 3)    & (0.80 , 0.57)                 & (0.80 , 0.57)                            & \textbf{(0.93 , 0.86)}  \\
MMSKCC & (0, 3)     & (0 , 0.80)                 & (0 , 0.50)                            & \textbf{(0 , 1)}  \\
Hospital with one WSI           & (3, 5)     & \textbf{(0 , 1)}                & (0.75 , 0.75)                         & \textbf{(0 , 1)}  \\ \hline
\textbf{Overall}                   &  (34, 40)                     & 71.60\%                     & 71.60\%          & \textbf{79.70\%}\\ \hline
\end{tabular}%
% }

\label{exteranlf1braintable}
\end{table*}

\subsection{Bias Reduction via Ranking Loss}
As mentioned previously, the main motivation behind the proposing RFL came from the  ability of the extracted features based on primary diagnosis labels to classify the hospitals. Therefore, using the proposed RFL, it is expected to reduce the accuracy of institution (hospital) classification since for each image, we sequestered the other images of same hospital to help model.  In other words,  we prevented  the model to learn the specific pattern from hospital identifiers. Consequently, the accuracy of hospital classification should be decreased. To investigate this effect,  %we trained three networks, Classification, RLL, and SRLL, for all images in the training dataset. After training the networks, we put models in evaluation mode, and then we extracted a feature vector for each image. 
the feature extracted using three mentioned models in previous experiments (i.e., classification, RFL, and ISL models) which are obtained based on primary diagnosis,  labels are utilized  to classify the institutes. The accuracy of the classifying the institutions shows the ability of  three mentioned models to learn hospital features in addition to cancer-related  features during the training phase.
For this purpose, the  Support Vector Machine (SVM) model with Radial Basis Function (RBF) as kernel was employed. Finally, we utilized the test images feature vectors to evaluate the SVM classifiers.

Table~\ref{svmhospitaltable} shows the accuracy of hospital classification for different models on two cancer types. %The results show that the feature vectors extracted from the classification model to classify two subtypes of the Brain and Lung dataset can  classify $32$ and $40$ hospitals, respectively. However, the SVM models for RLL and SRLL could not obtain good accuracy.
As it is reported,  the extracted feature vectors from the classification model suffer from hospital bias, since it learned to distinguish source site institutions. 
The RFL model reduced the accuracy of hospital classification dramatically and ISL model reduced it further. The dropping accuracy to 30\% (from 73\% to 41\%) for the Lung dataset and  to 25\% (from 68\% to 43\%) for Brain dataset in the ISL model shows that the ranking loss function reduces bias in the trained model to learn the source site institution. In essence, instead of learning source site institution during the training, the sequestering with ranking loss model concentrates on salient hospital-agnostic image features.

% \usepackage{multirow}
\begin{table}[]
\centering
\caption{A comparison  of the F1-scores for  hospital classification. The features extracted by  ISL function has the lowest hospital classification accuracy that shows the bias has been reduced (\textbf{EN=EfficientNet}).}
% \scalebox{1.05}{
\begin{tabular}{|c|c|c|c|}
\hline
\cellcolor[HTML]{EFEFEF}            & \cellcolor[HTML]{EFEFEF}      & \cellcolor[HTML]{EFEFEF}   & \cellcolor[HTML]{EFEFEF}      \\
 \multirow{-2}{*}{\cellcolor[HTML]{EFEFEF}\textbf{Dataset}} & \multirow{-2}{*}{\cellcolor[HTML]{EFEFEF}\textbf{EN}} & \multirow{-2}{*}{\cellcolor[HTML]{EFEFEF}\textbf{RLF}} & \multirow{-2}{*}{\cellcolor[HTML]{EFEFEF}\textbf{ISL}} \\ \hline
 Lung Cancer                                                       & 73.2\%                                                              & 45.9\%                                                   & \textbf{41.0\%}                                           \\  \hline Brain Cancer                                                     & 68.9\%                                                              & 51.7\%                                                   & \textbf{43.8\%}                                           \\ \hline
\end{tabular}
\label{svmhospitaltable}
\end{table}
%####################################
\section{Concluding Remarks}
\label{Conclusions}
Recent progress in deep neural networks in pathology has led to the accelerated adoption of digital pathology. This development creates opportunities o devise novel solutions for various problems in anatomic pathology. However, due to challenges such as the large WSI sizes and implicit bias in image datasets,  designing and implementing a computer-aided diagnosis system faces some serious obstacles.   To address these challenges and train a DNN-based customized for the image search and reduce the bias effect on the results, we proposed a novel ranking loss function (RFL). The new loss function enables us to train a network based on image search. In addition, we proposed instance sequestering learning (ISL) to alleviate the bias during the training using isolating the given hospital-specific patterns.

 The proposed models were compared with two state-of-the-art DNNs. The results indicate that the trained model is more accurate in image search of Lung and Brain tumour types. Furthermore, external validation was conducted by the alternate exclusion of hospitals during the training phase to assess the proposed loss function in reducing bias. A comparison of results obtained using the classification and ISL clearly shows that the accuracy of image search is increased on images from the hospitals that have not contributed to the training, and the bias is significantly reduced.

\vspace{0.1in}
\textbf{Acknowledgement} This research was supported in part by a grant from Ontario Research Funds - Research Excellence. As well, the work was supported by a start-up grant from Mayo Clinic (for H.R.Tizhoosh) and NSERC Discovery Grants (for S.Rahnamayan and H.R.Tizhoosh).

%%%%%%%%% REFERENCES
{
    \small
    \bibliographystyle{ieeetr}
    \bibliography{macros, main}
}

\end{document}